\documentclass[12pt,english]{article}
\usepackage{
amsmath,
amssymb,
amsthm,
amsxtra,
authblk,
babel,
braket,
booktabs,
caption,
cite,
float, 
graphbox,
graphicx,
hyperref,
dsfont,
pgfplots,
siunitx,
subfig,
subfloat,
tabu,
xcolor,
xparse,
}

\pgfplotsset{compat=1.14}
\hypersetup{colorlinks=true,
	linkcolor={red!30!black},
    	citecolor={blue!50!black},
    	urlcolor={blue!30!black}
   	 }

\newcommand{\eqb}{\begin{equation}}
\newcommand{\eqe}{\end{equation}}
\newcommand{\dmb}{\begin{displaymath}}
\newcommand{\dme}{\end{displaymath}}

\newcommand{\eab}{\begin{eqnarray}}
\newcommand{\eae}{\end{eqnarray}}

\newcommand{\be}{\begin{equation}}
\newcommand{\ee}{\end{equation}}

\RenewDocumentCommand\[{}{\begin{equation}}
\RenewDocumentCommand\]{}{\end{equation}}
\NewDocumentCommand\der{}{\mathrm{d}}

\NewDocumentCommand\intkern{}{\int\kern-5pt}

\NewDocumentCommand\lagr{}{\mathcal{L}}

\NewDocumentCommand\projl{}{(\mathds{1}-\gamma^5)}
\NewDocumentCommand\projr{}{(\mathds{1}+\gamma^5)}
\NewDocumentCommand\rpm{}{\raisebox{.195ex}{$\pm$}\hspace{0.05em}}

\usepgfplotslibrary{groupplots}

\title{\bf General Neutrino Interactions at the DUNE Near Detector}
\author{Ingolf Bischer, Werner Rodejohann}

\date{{\it 
Max-Planck-Institut f\"ur Kernphysik, \\Postfach
103980, D-69029 Heidelberg, Germany}\\
\href{mailto:bischer@mpi-hd.mpg.de}{bischer@mpi-hd.mpg.de}, 
\href{mailto:werner.rodejohann@mpi-hd.mpg.de}{werner.rodejohann@mpi-hd.mpg.de}
\\[2ex]%
}

\begin{document}
  \maketitle
\begin{abstract}
\noindent
We consider the effect of general neutrino interactions (scalar, vector, pseudoscalar, axial vector and tensor) in neutrino-electron scattering at the DUNE near detector. Those interactions can be associated with heavy new physics and their effect is to cause distortions in the recoil spectrum of the electrons. We show that for some cases energy scales up to 9 TeV are accessible after a 5 year run and that current bounds on interaction parameters can be improved by up to an order of magnitude. The full set of general interactions includes the usually considered 
neutrino-electron non-standard matter interactions, and the near detector will give limits comparable 
but complementary to the ones from the analysis of neutrino oscillations in the far detector. 
\end{abstract}
\newpage

\section{Introduction}
As neutrino physics parameters are pinned down with ever more precision, windows to potential new physics are opened. The phenomenologically interesting neutrino non-standard interactions (NSI) (see e.g.\ \cite{nsireview1,nsireview2} for reviews) constitute a simple, model-independent and experimentally accessible parametrisation of new physics associated with some high energy scale or with a small coupling to the Standard Model of particle physics (SM). In the framework of effective dimension-six operators, one can, however, consider a more general set of operators involving (pseudo)scalar and tensor currents. Unlike NSI terms, such interactions do not leave a measurable effect when neutrinos propagate in matter \cite{Bergmann:1999rz}. In addition, their properties depend, for instance, on the Dirac or Majorana nature of neutrinos. If found, such exotic new interactions could in fact help resolve this fundamental question \cite{Rosen:1982pj,xu}. 
In this paper we will consider the effects of new neutrino interactions (scalar, vector, pseudoscalar, axial vector and tensor) in neutrino-electron scattering. Those exotic interactions change not only the total cross section but also the recoil spectrum of the scattered electrons. 

A potent future experiment to address precision measurements in neutrino physics is the Deep Underground Neutrino Experiment (DUNE) which will observe neutrino oscillations at a \SI{1300}{km} long baseline \cite{design-report-1}. 
In this work, we are not mainly concerned with the primary physics goals of the experiment like determining the leptonic charge-parity (CP) violating phase and the neutrino mass ordering. Instead, we focus on the possibilities that the high-intensity neutrino beam produced for DUNE at the Long-Baseline Neutrino Facility (LBNF) opens up for non-oscillation physics.\footnote{Other future possibilities to probe neutrino-electron scattering also exist, for instance IsoDAR \cite{Conrad:2013sqa}, and the nu\-STORM facility \cite{Long:2018zpc}.} Indeed, the planned near detector (ND) located \SI{575}{m} away from the target provides the opportunity for a physics programme of its own. In this article, as one aspect of ND physics, we investigate neutrino-electron scattering events. This has been previously considered in \cite{falkowski}, discussing constraints on NSI from total cross-sections and their embedding in the SM effective field theory (SMEFT) framework. Here we provide an extended analysis which distinguishes itself firstly by the consideration of general new interactions and secondly by employing differential cross-sections to make full use of the tentative detector capabilities. This framework leads to improved (and several new) bounds while being able to discriminate the signatures of different types of new interactions and suitable for an immediate inclusion of specific experimental uncertainties. 
In our framework, electron NSI are included as a special case, and hence 
this procedure is independent of and in parts complementary to the studies of NSI in oscillation 
physics at DUNE \cite{deGouvea:2015ndi,coloma,blennow,Deepthi:2017gxg}, where typically not electron NSI but matter NSI 
(including the appropriate linear combination of electron, up and down quark interactions) are measurable. In fact the limits on electron NSI are comparable in the near and far detector analyses. 
\\

This document is structured as follows. In \autoref{sec:nsi}, our parametrisation of general neutrino-electron interactions beyond the SM is reviewed. In \autoref{sec:dune}, we then discuss the experimental setup and relevant cross sections for DUNE ND electron neutrino scattering and derive predictions for future bounds on new interactions before concluding in \autoref{sec:conclusion}. The origin of the new interactions from gauge invariant dimension-6 operators, a mapping between different parametrisations, and technical details on the derivation of the cross section are delegated to the appendix. 

\section{General neutrino-electron interactions}
\label{sec:nsi}
The Lagrangian of the most general $\nu\nu ee$ effective interaction, containing 
scalar,   vector, pseudoscalar, axial vector and tensor terms, can be parametrised in terms of Fermi's constant $G_F$ by
\[\label{eq:lagrangian}
\lagr=-\frac{G_F}{\sqrt{2}}\sum_{\alpha,\beta}\sum_{j=1}^{10}\left(\stackrel{\phantom{j}(\sim)j}{\epsilon}\right)_{\alpha\beta}
\left(\overline{\nu}^\alpha\mathcal{O}_j\nu^\beta\right)
\left(\overline{e}\mathcal{O}_j'e\right),
\]
where $\alpha$ and $\beta$ label flavors, and the operators $\mathcal{O}_j,\mathcal{O}_j'$ and $\epsilon$-parameters are listed in \autoref{tab:chiral-operators}. The possible origin of this Lagrangian from gauge-invariant dimension-6 operators consisting of SM particles is discussed in \autoref{sec:smeft}.   
Note the presence of the left- and right-projection operators in the definition of the operators in \autoref{tab:chiral-operators}. Thus, terms parametrised by $\widetilde\epsilon^j$ and generally all exotic interactions ($j\geq5$) require right-handed neutrinos (either charge-conjugates of $\nu_L^\alpha$ in the case of Majorana neutrinos or their right-handed counterpart $N_R^\alpha$ in the case of Dirac neutrinos). In this paper we will not discuss any effects of potentially heavy sterile neutrinos, but only consider right-handed counterparts of the active neutrinos whose masses can be neglected at the considered energy scales. 
In the Standard Model (SM), the only non-vanishing coefficients are $\epsilon^L$ and $\epsilon^R$ which receive (flavor-diagonal) contributions from neutral-current and (after Fierz transformation) charged-current interactions in Fermi theory. 
That is, the SM Lagrangian is obtained from Eq.\ (\ref{eq:lagrangian}) by setting all $\epsilon$ to zero, except for 
\[\label{eq:gsm}
\epsilon^{L,\mathrm{SM}}_{\alpha\beta}= \delta_{\alpha e}\delta_{\beta e} + \left(-\frac12+s_W^2\right)\delta_{\alpha\beta}\,,
\qquad
\epsilon^{R,\mathrm{SM}}_{\alpha\beta}= s_W^2\delta_{\alpha\beta}\,, 
\]
where $s_W$ denotes the sine of the Weinberg angle, and the first term in $\epsilon^{L,\mathrm{SM}}$ is the charged-current contribution. The framework presented here is more general than the typically investigated non-standard interactions (NSI) which correspond to $\epsilon^L$ and $\epsilon^R$ or their linear combinations $\epsilon^V=\epsilon^L+\epsilon^R$ and $\epsilon^A=\epsilon^L-\epsilon^R$, the former being of major interest when studying neutrino oscillations in matter. Without loss of generality, one can omit adding a complex conjugate in \eqref{eq:lagrangian} and instead impose
\[
\begin{split}
\epsilon^L_{\alpha\beta}=\epsilon^{L*}_{\beta\alpha}\,,\qquad
\widetilde\epsilon^L_{\alpha\beta}=\widetilde\epsilon^{L*}_{\beta\alpha}\,,\quad & \quad
\epsilon^R_{\alpha\beta}=\epsilon^{R*}_{\beta\alpha}\,,\qquad
\widetilde\epsilon^R_{\alpha\beta}=\widetilde\epsilon^{R*}_{\beta\alpha}\,,\\
\epsilon^{S}_{\alpha\beta}=\widetilde\epsilon^{S*}_{\beta\alpha}\,,\qquad
\epsilon^{P}_{\alpha\beta}&=-\widetilde\epsilon^{P*}_{\beta\alpha}\,,\qquad
\epsilon^{T}_{\alpha\beta}=\widetilde\epsilon^{T*}_{\beta\alpha}\,.
\label{eq:epsilonconstraints}
\end{split}
\]
In particular, the parameters in the first line, which are associated with NSI (and their counterpart involving right-handed neutrinos), are hermitian matrices in flavor space and thus real in their flavor-diagonal entries. 
The parametrisation in Eq.\ (\ref{eq:lagrangian}) is useful when dealing with chiral particles, as for instance left-handed neutrinos and right-handed antineutrinos produced from the pion decays at LBNF. For completeness, we mention that an equivalent parametrisation has been employed in parts of the literature, e.g.\ \cite{Kayser:1979mj,Rosen:1982pj,xu}; we give the mapping between the two parametrisations in \autoref{sec:CDparametrisation}. 
\begin{table}
{\centering
\begin{tabu}spread 50pt{X[$,c]X[$,c]X[$,c]X[$,c]}
\toprule
j & \stackrel{(\sim)}{\epsilon_j} & \mathcal{O}_j & \mathcal{O}_j'\\
\midrule
1&\epsilon_L & 		\gamma_\mu \projl & \gamma^\mu \projl\\
2&\tilde\epsilon_L &  \gamma_\mu \projr & \gamma^\mu \projl\\
3&\epsilon_R &		\gamma_\mu \projl & \gamma^\mu \projr\\
4&\tilde\epsilon_R &  \gamma_\mu \projr & \gamma^\mu \projr\\
5&\epsilon_S &		\projl & \mathds{1} \\
6&\tilde\epsilon_S &  \projr & \mathds{1} \\
7&-\epsilon_P &		\projl & \gamma^5 \\
8&-\tilde\epsilon_P & \projr & \gamma^5 \\
9&\epsilon_T &		\sigma_{\mu\nu} \projl & \sigma^{\mu\nu} \projl\\
10&\tilde\epsilon_T &  \sigma_{\mu\nu} \projr & \sigma^{\mu\nu} \projr\\
\bottomrule
\end{tabu}
\caption{Coupling constants and operators appearing in the general $\nu\nu ee$ interaction Lagrangian \eqref{eq:lagrangian}.}
\label{tab:chiral-operators}
}
\end{table}

In the case of Dirac neutrinos, all the above parameters are a priori unconstrained. We note two interesting special cases. Firstly, imposing CP-invariance would require that all parameters be real. 
Secondly, as has been previously discussed \cite{Rosen:1982pj,xu}, if neutrinos are of Majorana nature, there are also more constraints on these parameters. In our parametrisation and for general flavor combinations, it is straightforward to show that Majorana nature of neutrinos implies
\[\label{eq:majoranacondition}
\epsilon^L_{\alpha\beta} = - \widetilde\epsilon^L_{\beta\alpha}\,,\quad
\epsilon^R_{\alpha\beta} = - \widetilde\epsilon^R_{\beta\alpha}\,,\quad
\epsilon^S_{\alpha\beta} = \epsilon^S_{\beta\alpha}\,,\quad
\epsilon^P_{\alpha\beta} = \epsilon^P_{\beta\alpha}\,,\quad
\epsilon^T_{\alpha\beta} = -\epsilon^T_{\beta\alpha}\,.
\]
Consequently, the number of independent parameters is almost halved with respect to the Dirac case.\footnote{Considering all possible flavor combinations, in the general case there are 90 real degrees of freedom, which are reduced to 48 in the Majorana case. Further details on this statement are reserved for future work.} 

In this work we take an agnostic position towards mass models, CP-, and lepton flavor violation and hence consider all parameters. As  we will discuss later, all new physics signatures in neutrino-electron scattering at DUNE ND may arise from Majorana-allowed new interaction parameters, such that no light can be shed on the question of Dirac or Majorana nature. This is not in conflict with the results from \cite{Rosen:1982pj} and \cite{xu}, since their generalised interactions are lepton flavor conserving. Likewise, all observables may be generated from CP-conserving interactions.
\\

It is common to translate the $\epsilon$-parameters into mass scales $M/g'$ that would be associated with a heavy particle of mass $M$ coupling with strength $g'$ to the leptons inducing the effective interactions at low energies. Namely, according to $\epsilon\,G_F/\sqrt{2}=g'^2/M^2$, one obtains the relation
\[\label{eq:massscale}
\frac{M}{g'}\approx\sqrt{\frac{\sqrt{2}}{\epsilon\,G_F}}\,.
\]
For the reader's convenience, we will translate via Eq.\ (\ref{eq:massscale}) our bounds on $\epsilon$-parameters to mass scales when presenting bounds in the following sections. 

\section{Neutrino-electron scattering at DUNE ND}
\label{sec:dune}
\subsection{Experimental specifications}
\label{sec:exp}
\begin{figure}
\centering
\includegraphics{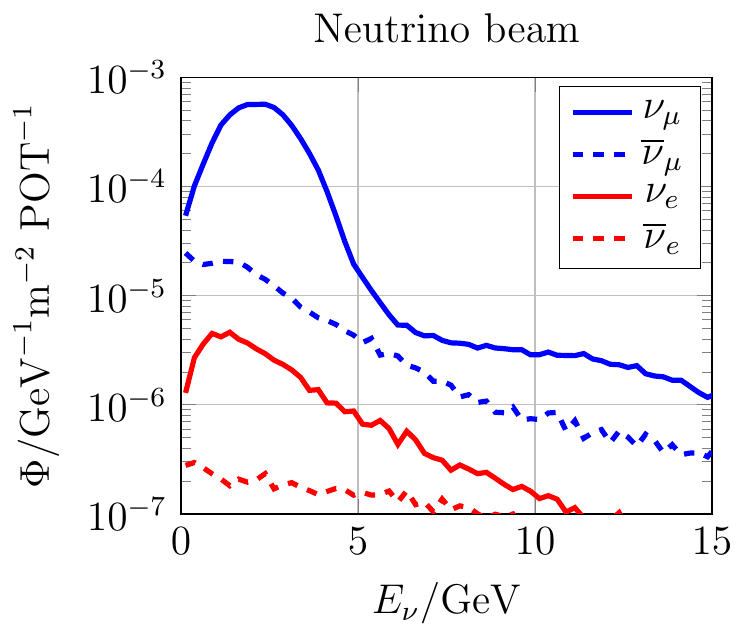}
\includegraphics{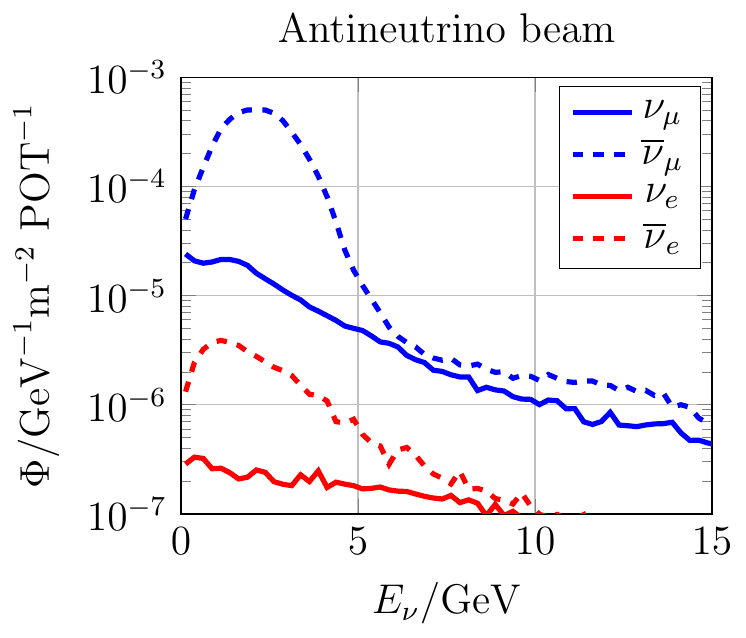}
\caption{Simulated neutrino fluxes from the optimised design in \cite{fluxes} for both neutrino and antineutrino beam. }
\label{fig:fluxes}
\end{figure}
To investigate the sensitivity of DUNE ND towards new physics effects in neutrino-electron scattering one needs to discuss both the neutrino beam and the ND design. Neutrino beams can be produced in two channels, neutrino and antineutrino. The composition of the two beam channels is shown in \autoref{fig:fluxes}, where we plot the simulated neutrino fluxes from \cite{fluxes}. We first note that the composition in the neutrino channel is approximately  90-95\% muon neutrinos, 5-10\% muon antineutrinos and about 1\% electron (anti)neutrinos, while in the antineutrino channel, the fractions of muon neutrinos and antineutrinos are reversed. In our calculation we neglect the small contamination by electron (anti)neutrinos for simplicity. Since neutrinos with energies above \SI{10}{GeV} are at least two orders of magnitude less abundant in the beam, we cut off the fluxes at this energy. The considered neutrino energy range is then $0.125$-$10$ GeV. In the first phase of operation, which is tentatively 5 years, a \SI{1.2}{MW} beam of \SI{120}{GeV} protons will be produced \cite{design-report-1} (the power is expected to double afterwards).  Taking 2.5 years of operation in each channel and assuming \num{1.8e7} seconds per year as the operational duration, this amounts to \SI{1.123e21}{POT\per a} (protons-on-target per year) or in total \SI{2.809e21}{POT} per channel in the first five years.

The ND conceptual design is yet undetermined, however, from recent reports \cite{neutrinoposter} we extract that the DUNE ND will presumably involve a liquid argon time projection chamber (LArTPC) with 84 tons of fiducial Argon mass. This amounts to \num{1.2663e30} atoms, such that we estimate the number of target electrons by $18\cdot\num{1.2663e30}=\num{2.2793e31}$. The energy resolution of the electromagnetic calorimeters is estimated as $0.06/\sqrt{E_e}\approx0.06/\sqrt{T}$ (in GeV) at least up to energies of \SI{10}{GeV} (which is our region of interest) \cite{design-report-4}.

\subsection{Differential cross sections and parametrisation}
In this section we discuss the relevant differential cross sections. In principle, when considering $\nu_\mu$-$e$ scattering, the final states may either be $\nu+e$ or $\nu+\mu$. However, we are neglecting the possibility of creating a muon instead of an electron as the charged final state because the threshold energy for muon production from neutrino-electron scattering is $m_\mu^2/2m_e\approx\SI{10.9}{GeV}$ which is above our considered energy range. Thus the only relevant (anti)neutrino-electron scattering processes are $\nu_\mu+e^-\rightarrow\nu_\beta+e^-$ and $\overline\nu_\mu+e^-\rightarrow\overline\nu_\beta+e^-$, where $\alpha$ denotes any flavor, while the SM only contributes to $\beta=\mu$. Since the flavors of the final neutrinos cannot be measured, one has to sum over all final flavors $\alpha$ which in turn implies that we are only sensitive to parameter sums.

The differential cross sections for $\nu_\mu+e\rightarrow\nu_\beta+e$ scattering and $\overline\nu_\mu+e\rightarrow\overline\nu_\beta+e$ scattering respectively read
\[
\begin{split} 
\frac{\der\sigma_{\nu_\mu\rightarrow\nu_\beta}}{\der T} &=
\frac{G_F^2m_e}{\pi}
\left[
A+2B\left(1-\frac{T}{E_\nu}\right)+C\left(1-\frac{T}{E_\nu}\right)^2+D\frac{m_eT}{E_\nu^2}
\right],\label{eq:diffcross}\\
\frac{\der\sigma_{\overline\nu_\mu\rightarrow\overline\nu_\beta}}{\der T} &=
\frac{G_F^2 m_e}{\pi}
\left[
C+2B\left(1-\frac{T}{E_\nu}\right)+A\left(1-\frac{T}{E_\nu}\right)^2+D\frac{m_eT}{E_\nu^2}
\right],
\end{split}
\]
where $E_\nu\gg m_e$ denotes the energy of the incoming (anti-)neutrino, $T<E_\nu$ the kinetic energy of the recoiled electron, and the coefficients are given by
\[\label{eq:abcd}
\begin{split}
A&=2|\epsilon^L_{\mu\beta}|^2 +\frac14(|\epsilon^S_{\mu\beta}|^2+|\epsilon^P_{\mu\beta}|^2) + 8|\epsilon^T_{\mu\beta}|^2 -2\mathrm{Re}\left((\epsilon^S+\epsilon^P)_{\mu\beta}\epsilon^{T*}_{\mu\beta}\right)
\,,\\
B&=-\frac14(|\epsilon^S_{\mu\beta}|^2+|\epsilon^P_{\mu\beta}|^2)+8|\epsilon^T_{\mu\beta}|^2
\,,\\
C&= 2|\epsilon^R_{\mu\beta}|^2 +\frac14(|\epsilon^S_{\mu\beta}|^2+|\epsilon^P_{\mu\beta}|^2) + 8|\epsilon^T_{\mu\beta}|^2 +2\mathrm{Re}\left((\epsilon^S+\epsilon^P)_{\mu\beta}\epsilon^{T*}_{\mu\beta}\right)
\,,\\
D&= -2\mathrm{Re}\left(\epsilon^L_{\mu\beta}\epsilon^{R*}_{\mu\beta}\right)+\frac12|\epsilon^S_{\mu\beta}|^2-8|\epsilon^T_{\mu\beta}|^2\,.
\end{split}
\]
We discuss the derivation (using \verb|Package-X| \cite{packagex}) in \autoref{sec:derivation}. 
Notice that apart from the $S/P$-$T$-mixed terms in $A$ and $C$, and the $L$-$R$-mixed term in $D$, all parameters appear in the form of their modulus squared. Furthermore, it is noteworthy that the contribution of $D$ is suppressed by a factor of $m_e/E_\nu\sim10^{-4}$ with respect to the contributions from $A$, $B$, and $C$. Hence it is obvious that, since the only difference between the contributions of scalar and pseudoscalar $\epsilon$-parameters is the appearance of the former in $D$, they will be practically indistinguishable.
In the SM case, the cross section coefficients simplify to
\[\label{eq:abcdSM}
A_{\mathrm{SM}} = 2 g_L^2\,,\quad B_{\mathrm{SM}}=0\,,\quad C_{\mathrm{SM}}= 2g_R^2\,,\quad D_{\mathrm{SM}} =-2g_Lg_R\,,
\]
where $g_L=-1/2+s_W^2$ and $g_R=s_W^2$. Throughout the computations we use the value $s_W^2=0.22343$ (which is obtained in the on-shell scheme) \cite{pdg2018}.

Let us discuss, which parameters are accessible independently after summing over all final neutrino flavors. The exotic interactions $S$, $P$, and $T$ are straightforward: Apart from the mixed term involving (pseudo)scalar and tensor parameters which can be neglected when considering single-parameter extensions (i.e.\ one new parameter at a time), the setup is only sensitive to the sum of square-moduli, for which we define the effective parameter
\[\label{eq:epseffective}
|\epsilon_\mu^j|^2\equiv\sum_{\beta=e,\mu,\tau}|\epsilon_{\mu\beta}^j|^2\,, \quad j=S,P,T\,.
\]
Since there is a sum of squares on the right-hand side of \eqref{eq:epseffective}, bounds on $\epsilon^j_\mu$ equivalently translate to (conservative) bounds on $|\epsilon^j_{\mu e}|$,  $|\epsilon^j_{\mu \mu}|$, and $|\epsilon^j_{\mu \tau}|$.
Concerning the NSI-type interaction, we split $\epsilon^L$ and $\epsilon^R$ into SM plus new physics contributions,
\[
\epsilon^L_{\mu\beta} = \epsilon^{L,\mathrm{SM}}_{\mu\beta} + \epsilon^{L,\mathrm{NSI}}_{\mu\beta}\,,\quad \epsilon^R_{\mu\beta} = \epsilon^{R,\mathrm{SM}}_{\mu\beta} + \epsilon^{R,\mathrm{NSI}}_{\mu\beta}\,,
\]
where the SM expressions are given in Eq.\ (\ref{eq:gsm}).
Hence, the effective parameter introduced in \eqref{eq:epseffective} would become
\[\label{eq:lrsplit}
\sum_{\beta}\left|g_j\delta_{\mu\beta}+\epsilon^{j,\mathrm{NSI}}_{\mu\beta}\right|^2 = \left(g_j+\epsilon^{j,\mathrm{NSI}}_{\mu\mu}\right)^2+|\epsilon^{j,\mathrm{NSI}}_{\mu}|^2\,, \quad j=L,R\,,
\]
where we used the fact that $\epsilon^{j,\mathrm{NSI}}_{\mu\mu}$ is real and defined a different effective parameter, namely excluding the flavor-diagonal contribution,
\[\label{eq:epsRLeffective}
|\epsilon_\mu^{j,\mathrm{NSI}}|^2\equiv\sum_{\beta=e,\tau}|\epsilon_{\mu\beta}^j|^2\,, \quad j=L,R\,.
\]
Since only the flavor-diagonal NSI parameters interfere with the SM in \eqref{eq:lrsplit}, we deduce that they will be more constrained than the off-diagonal parameters
and that we have to investigate them separately. In principle, the product of $\epsilon^L_{\mu\beta}$ and $\epsilon^{R*}_{\mu\beta}$ in $D$ would contain further superpositions of $L$- and $R$-NSI parameters, when tuning both beyond the SM,
\[\label{eq:lrmixed}
\sum_\beta \mathrm{Re}\left(\epsilon^L_{\alpha\beta}\epsilon^{R*}_{\alpha\beta}\right) =
g_Lg_R+g_L\epsilon^{R,\mathrm{NSI}}_{\alpha\alpha}+g_R\epsilon^{L,\mathrm{NSI}}_{\alpha\alpha} + \sum_\beta \mathrm{Re}\left(\epsilon^{L,\mathrm{NSI}}_{\mu\beta}\epsilon^{R,\mathrm{NSI}*}_{\mu\beta}\right).
\]
However, since this term appears in $D$, which is suppressed by $m_e/E_\nu\sim10^{-4}$, it is reasonable to consider only flavor-diagonal contributions or only the SM part in \eqref{eq:lrmixed}, to avoid more complicated dependencies like in rightmost term which constitutes subleading new physics effects. Practically that means that we set this term to zero in our calculations unless we consider flavor diagonal $R$ \emph{and} $L$ extensions, such that
\[
\sum_\beta \mathrm{Re}\left(\epsilon^L_{\alpha\beta}\epsilon^{R*}_{\alpha\beta}\right) =
g_Lg_R+g_L\epsilon^{R,\mathrm{NSI}}_{\alpha\alpha}+g_R\epsilon^{L,\mathrm{NSI}}_{\alpha\alpha} + \epsilon^{L,\mathrm{NSI}}_{\mu\mu}\epsilon^{R,\mathrm{NSI}}_{\mu\mu}\,.
\]
In conclusion, we expect to obtain bounds on the parameters listed in \autoref{tab:parameters}. As previously stated, all of the observables may originate from parameters which are allowed in the case of Majorana neutrinos unless we assume flavor conservation, such that $|\epsilon^T_{\alpha}| = |\epsilon^T_{\alpha\alpha}| = 0$.


\begin{table}
\begin{tabu}{X[1.2]X[$,c]X[$,2.2,c]X[$,3,c]}
\toprule
Observable		&\epsilon^{L/R,\mathrm{NSI}}_{\mu\mu} 
				&|\epsilon^{L/R,\mathrm{NSI}}_{\mu}| 
				&|\epsilon^{S/P/T}_{\mu}| 
				\\
\addlinespace[10pt]
Bound on	&	\epsilon^{L/R,\mathrm{NSI}}_{\mu\mu}
					& 	|\epsilon^{L/R,\mathrm{NSI}}_{\mu e}|\,,\; 
					|\epsilon^{L/R,\mathrm{NSI}}_{\mu\tau}|
					& 	|\epsilon^{S/P/T}_{\mu e}|\,,\;
					|\epsilon^{S/P/T}_{\mu\mu}|\,,\;
					|\epsilon^{S/P/T}_{\mu \tau}|
				\\
\bottomrule
\end{tabu}
\caption{Parameters that influence the scattering of electrons with muon (anti)neutrinos  
(first line) and the fundamental new interaction parameters that can be constrained from the respective measurements 
(second line).}
\label{tab:parameters}
\end{table}

\subsection{Spectra and sensitivities}
\begin{figure}
\centering{
\includegraphics{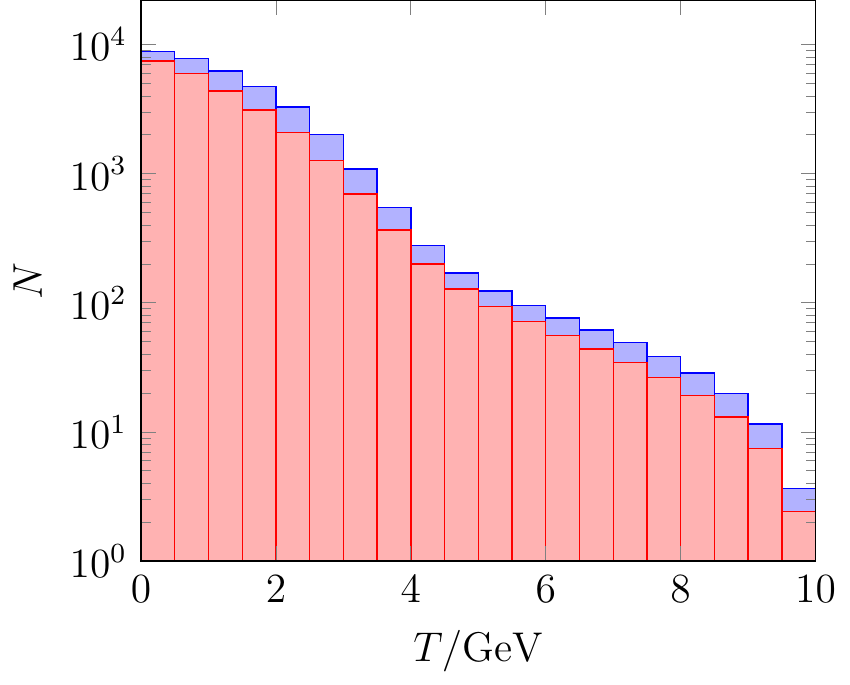}
\caption{Expected electron event numbers from neutrino-electron scattering in neutrino channel (blue) and antineutrino channel (red) at each 2.5 years of exposure assuming SM parameters.}
\label{fig:spectrum}
}
\end{figure}

\begin{figure}
\centering
\includegraphics[scale=1]{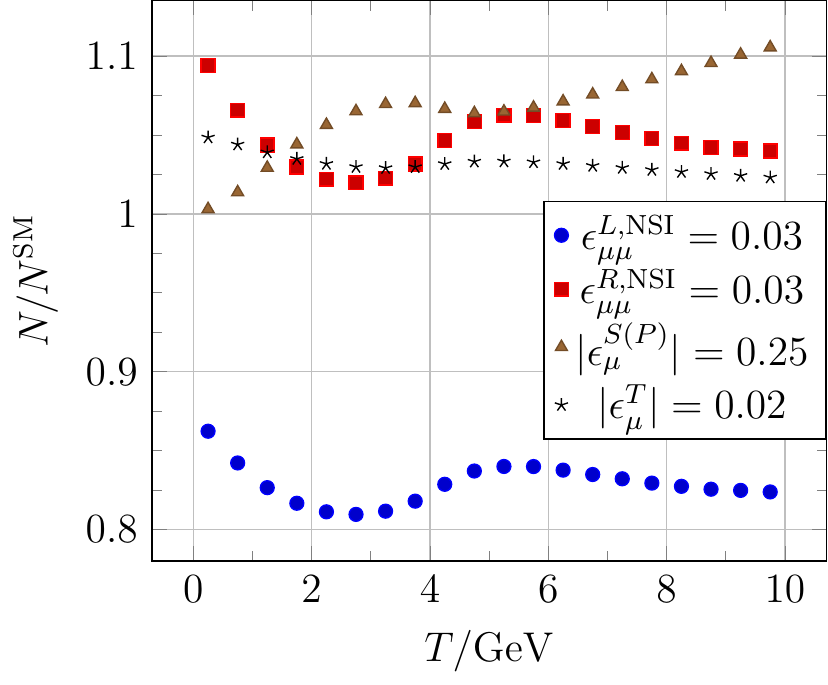}
\caption{Deviation of the expected event number spectrum in the neutrino channel for different new physics parameters tuned away from their SM values. The sample values are of the order of magnitude of current bounds, see \autoref{tab:bounds}.}
\label{fig:spectra}
\end{figure}

To make use of the spectral information, we consider event numbers within energy bins of the recoil electron of size $\Delta T$. The expected number of events in the $i$-th bin $[T_i, T_i+\Delta T]$ for a given type of neutrino $X=\nu_\mu,\overline\nu_\mu$ can be calculated as 
\[
N_i(\vec\epsilon) = \Delta t N_e \int_{0}^{E_{\rm max}}\der E_\nu \int_{\Delta T\cdot (i-1)}^{\Delta T\cdot i}\der T\,
\Theta(T-E_\nu)\frac{\der\sigma_X}{\der T}(E_\nu,T,\vec\epsilon)\frac{\der\Phi_X}{\der E_\nu} (E_\nu)\,,
\]
where $\Delta t$ is the time of data taking times the beam-specific POT (per time), $N_e$ is the number of electron targets, $E_{\rm max}=\SI{10}{GeV}$ is the maximal neutrino energy, $\Theta$ denotes the Heaviside function to assert the kinematic condition $T\leq E_\nu$, and $\der\Phi_X/\der E_\nu$ is the differential neutrino flux of type $X$ in units of neutrinos/GeV/m$^2$/POT.
We introduced the shorthand notation 
\[
\vec\epsilon=\left(\epsilon^{L,\mathrm{NSI}}_{\mu\mu},\,\epsilon^{R,\mathrm{NSI}}_{\mu\mu},\,
|\epsilon^{L,\mathrm{NSI}}_{\mu}|,\,|\epsilon^{R,\mathrm{NSI}}_{\mu}| ,\,
|\epsilon^{S}_{\mu}|,\,|\epsilon^{P}_{\mu}| ,\,|\epsilon^{T}_{\mu}| \right)
\]				
to make the dependence on all seven independent observables in \autoref{tab:parameters} explicit. To account for finite resolution effects, we convolve this spectrum with a Gaussian resolution function to predict realistic event numbers. Assuming SM parameter values and a bin size of $\Delta T=\SI{0.5}{GeV}$, we obtain the spectrum in \autoref{fig:spectrum}. 
The distinct effect of different single $\epsilon$-parameters tuned away from their SM values is demonstrated by \autoref{fig:spectra}, where we plot the associated spectra in relation to the SM prediction.

To project the sensitivities towards new physics, we apply a standard $\chi^2$-test, which seems adequate considering the sufficiently large event numbers in low-energy bins. Due to a lack of statistics at high energies, see \autoref{fig:spectra}, we use only the bins up to energies of $T=\SI{5}{GeV}$ (we have checked that the obtained bounds do not sensitively depend on the precise choice of this ``cutoff"). We employ the following $\chi^2$-function
\[
\chi^2(\vec\epsilon) = \frac{a^2}{\sigma_a^2}+\sum_{X=\nu_\mu,\overline\nu_\mu}\sum_{i=1}^{n_{\mathrm{bins}}}
\frac{\left((1+a)N_i^X(\vec\epsilon)-N_i^{X,\mathrm{SM}}\right)^2}{(\sigma_i^X)^2(\vec\epsilon)}\,,
\]
where $N_i^X$ denotes the number of events in the $i$-th bin of the channel $X$ (either neutrino or antineutrino), and $\sigma_i^X$ denotes the uncertainty, which we assume to be statistically dominated, 
\[
\sigma_i^X(\vec\epsilon)=\sqrt{N_i^X(\vec\epsilon)}\,.
\]
The small parameter $a$ is introduced to account for an overall normalisation uncertainty $\sigma_a$ which will include, in particular, flux uncertainties. We will consider three configurations,
\[\label{eq:configurations}
\begin{split}
&\text{A (``ideal'')}:\\
&\text{B (``optimistic'')}:\\
&\text{C (``conservative'')}:
\end{split}
\begin{split}
&\qquad (\sigma_a,T_\text{th})=(0,0)\,, \\
&\qquad (\sigma_a,T_\text{th})=(1\%,0)\,, \\
&\qquad (\sigma_a,T_\text{th})=(5\%,\SI{500}{MeV}) \,,
\end{split}
\]
where $T_\text{th}$ denotes the threshold energy for detecting a recoil electron of this process.


\subsection{Results}

\begin{table}
\tabulinesep=1.2mm
\begin{tabu}{X[$,1.5]X[$,2]X[c,1]X[$,c,2.5]X[$,c,2]X[$,c]}
\toprule
	\mathrm{Observable}		
	& \mathrm{NP\,Parameters} &Config. & \mathrm{Future\;DUNE}& \text{CHARM-II}& \frac{M}{g'}[\si{TeV}]\\
\midrule
\epsilon^{L,\mathrm{NSI}}_{\mu\mu}  & \epsilon^{L,\mathrm{NSI}}_{\mu\mu} 
			&A &  \rpm0.0014    &[-0.06,0.02]	&9.3 \\
			&&B & \rpm0.0028	 &				&6.7 \\
			&&C & \rpm0.0038    &				&5.7 \\
			
\midrule			
\epsilon^{L,\mathrm{NSI}}_{\mu} &|\epsilon^{L,\mathrm{NSI}}_{e\mu}|\,,\;
			|\epsilon^{L,\mathrm{NSI}}_{\mu\tau}|
			&A & 0.028	&  & 2.1\\
			&&B & 0.039	 &				&1.8 \\
			&&C & 0.046    &				&1.6 \\
\midrule
\epsilon^{R,\mathrm{NSI}}_{\mu\mu} & \epsilon^{R,\mathrm{NSI}}_{\mu\mu}
			&A & \rpm0.0017		&[-0.06,0.02]	& 8.6\\
			&&B & \rpm0.0027	 &				&6.8 \\
			&&C & \rpm0.0031    &				&6.3 \\
			
\midrule
\epsilon^{R,\mathrm{NSI}}_{\mu} &	|\epsilon^{R,\mathrm{NSI}}_{e\mu}|\,,\;	
			|\epsilon^{R,\mathrm{NSI}}_{\mu\tau}|
			&A & 0.027	&  & 2.1\\
			&&B & 0.035	 &				&1.9 \\
			&&C & 0.037    &				&1.8 \\
			
\midrule
\epsilon^{S}_{\mu} 	&|\epsilon^{S}_{e\mu}|\,,\;|\epsilon^{S}_{\mu\mu}|\,,\;
			|\epsilon^{S}_{\mu\tau}|
			&A &	0.10	& 0.4 & 1.1\\
			&&B & 0.12	 &				&1.0\\
			&&C & 0.14    &				&0.9 \\
			
\midrule
\epsilon^{P}_{\mu} 	&|\epsilon^{P}_{e\mu}|\,,\;|\epsilon^{P}_{\mu\mu}|\,,\;
			|\epsilon^{P}_{\mu\tau}|
			&A &	0.10	& 0.4 & 1.1\\
			&&B & 0.12	 &				&1.0\\
			&&C & 0.14    &				&0.9 \\
			
\midrule
\epsilon^{T}_{\mu} 	&|\epsilon^{T}_{e\mu}|\,,\;|\epsilon^{T}_{\mu\mu}|\,,\;
			|\epsilon^{T}_{\mu\tau}|
			&A &  0.008	& 0.04 & 4.0 \\
			&&B & 0.012	 &				&3.1\\
			&&C & 0.020    &				&2.4 \\
\bottomrule
\end{tabu}
\caption{Expected bounds (90\% CL) on new-physics neutrino-electron interaction parameters after 2.5+2.5 years of DUNE operation assuming three different experimental configurations (specified in \eqref{eq:configurations}), compared with current bounds from CHARM-II \cite{charm,charmb,xu,charmnsi}. For convenience, we list associated mass scales $M/g'$ according to \eqref{eq:massscale}.
}
\label{tab:bounds}
\end{table}

Concerning single-parameter bounds, namely, tuning only one parameter away from its SM value at a time we obtain the expected bounds shown in \autoref{tab:bounds}.
We conclude that the bounds on all of the general neutrino-electron scattering parameters will be improved after the first five years of DUNE operation. The frequently discussed NSI couplings can be constrained by roughly one order of magnitude, which is comparable to the results obtained in \cite{falkowski}. We will outline what our analysis adds to the discussion therein at the end of this section. 
Our bounds may also be compared to the expected bounds on matter NSI from oscillation data if we assume vanishing quark NSI.\footnote{Recall that matter NSI equal electron NSI if we assume quark NSI to vanish \cite{nsireview1}.} For propagation in matter only the vector part, $\epsilon^V=\epsilon^L+\epsilon^R$, is relevant. Assuming only one parameter to be non-zero at a time, we can extract from \autoref{tab:bounds} our weakest bound on parameters contributing to $\epsilon^V$ and compare this to the bounds from \cite{blennow}. Concerning flavor non-diagonal parameters we find
\[
\begin{split}
|\epsilon^V_{e\mu}|\leq 0.051\,,\quad
|\epsilon^V_{\mu\tau}|\leq 0.031\qquad\qquad\qquad  &\text{\cite{blennow},} \\
|\epsilon^{L,\mathrm{NSI}}_{e\mu}|,|\epsilon^{L,\mathrm{NSI}}_{\mu\tau}|,|\epsilon^{R,\mathrm{NSI}}_{e\mu}|,|\epsilon^{R,\mathrm{NSI}}_{\mu\tau}| \leq 0.039 \qquad\qquad &\text{\,\,this work.}
\end{split}
\]
Both methods lead to constraints at the same order of magnitude with differences in sensitivity to flavor combinations. Flavor-diagonal parameters cannot be measured independently in oscillations such that we can only compare our muon-flavored bounds to the following combinations,
\[
\begin{split}
\epsilon^V_{ee}-\epsilon^V_{\mu\mu}\in (-0.7,+0.8) \,,\quad
\epsilon^V_{\tau\tau}-\epsilon^V_{\mu\mu}\in (-0.08,+0.08)\qquad\qquad  &\text{\cite{blennow},} \\
|\epsilon^{L,\mathrm{NSI}}_{\mu\mu}|,|\epsilon^{R,\mathrm{NSI}}_{\mu\mu}| \in (-0.0027, 0.0027) \qquad\qquad\qquad &\text{\,\,this work,}
\end{split}
\]
such that in our case the muon flavor is strongly constrained, but no conclusions about other flavors can be drawn while oscillation constrains two flavor combinations, but less stringently.
Considering that constraints on neutrino-quark NSI are also expected to dramatically improve 
from neutrino-nucleus scattering at the ND \cite{falkowski} 
(as well as from coherent elastic neutrino-nucleus scattering \cite{Lindner:2016wff,Kosmas:2017tsq,AristizabalSierra:2018eqm})\footnote{For comparison, we note that the current bounds on general neutrino-quark interactions from the COHERENT experiment \cite{Akimov:2017ade} are at the order of $\epsilon\sim10^{-1}-10^{-2}$ neglecting potential degeneracies in parameter space \cite{Kosmas:2017tsq,AristizabalSierra:2018eqm}. }, 
we conclude that the determination of NSI parameters at 
the ND can serve as a valuable input to the fits of oscillation data. In particular, degeneracies between NSI and the CP phase \cite{Liao:2016hsa} may be lifted if the off-diagonal matter NSI parameters are constrained down to $\mathcal{O}(10^{-2})$ from an independent measurement \cite{Hyde:2018tqt}.
 \\

\begin{figure}
\centering{
\begin{tikzpicture} 
\node[anchor=south west,inner sep=0,align=center] (image) at (0,0) {%
\subfloat{\includegraphics[]{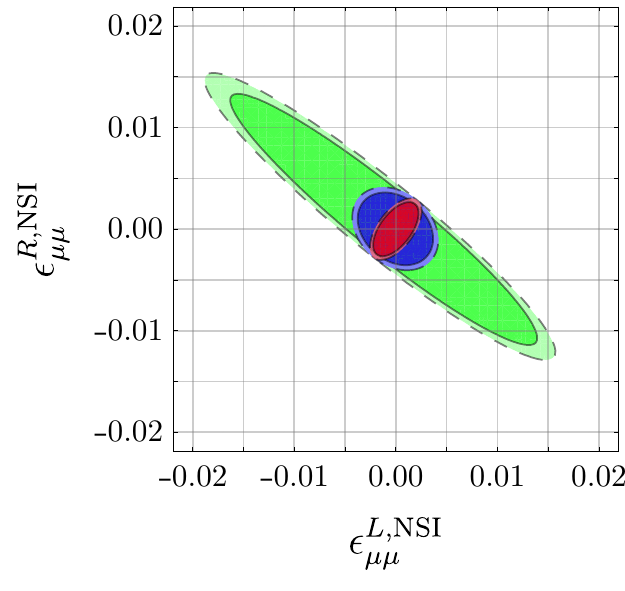}}\quad
\subfloat{\includegraphics[]{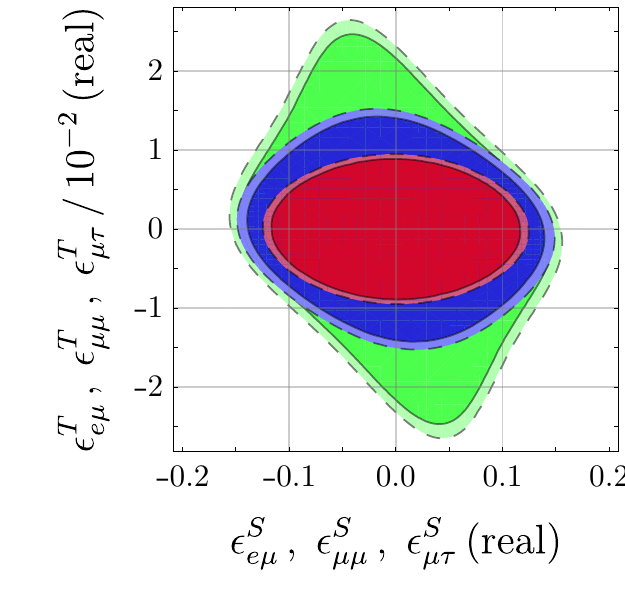}}\\
\subfloat{\includegraphics{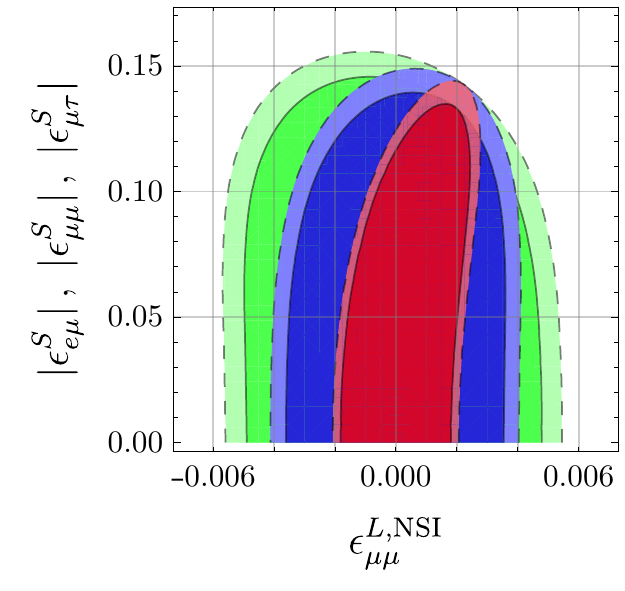}}\quad
\subfloat{\includegraphics{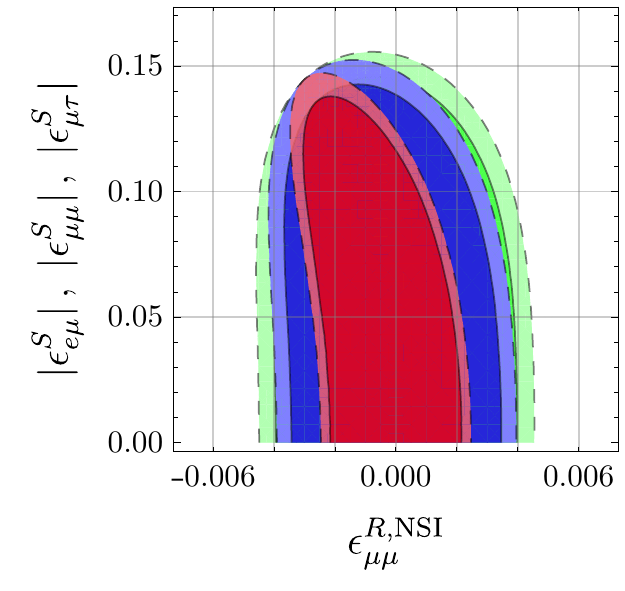}}\\
\subfloat{\includegraphics{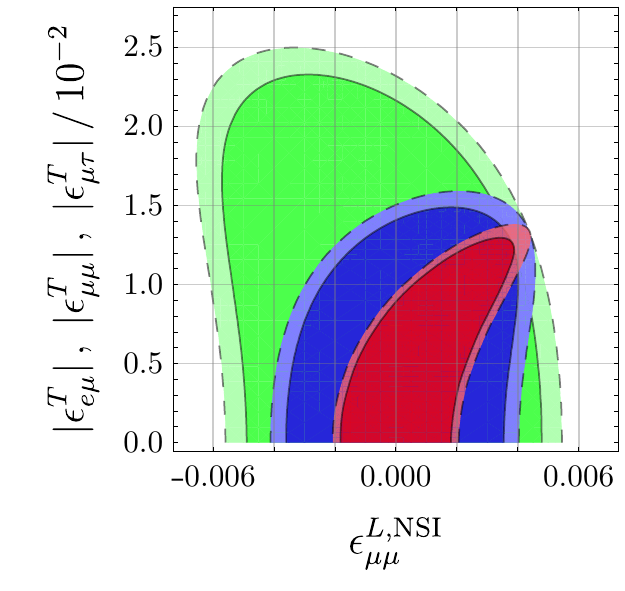}}\quad
\subfloat{\includegraphics{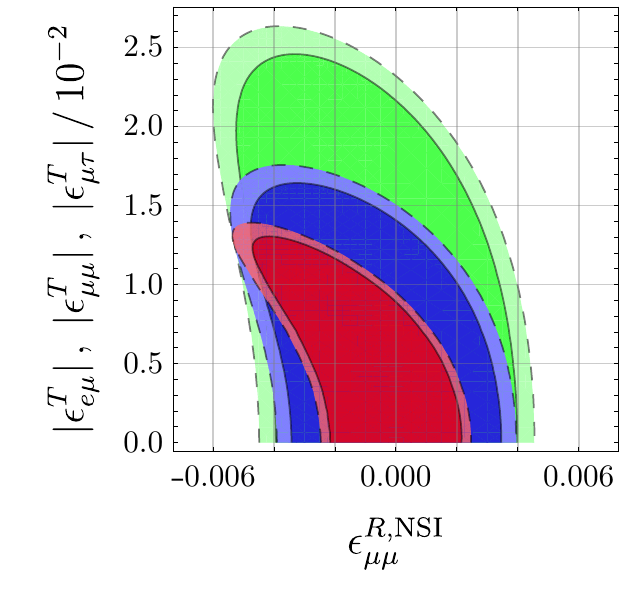}}
};
\end{tikzpicture}
}
\caption{Two-parameter exclusion plots after 2.5+2.5 years of DUNE operation. Contours correspond to 90\% and 95\% CL respectively. The red, blue, and green regions correspond to, respectively, experimental configurations A, B, and C (specified in \eqref{eq:configurations}).
We only show flavor-diagonal components of $\epsilon^{L,\mathrm{NSI}}$ and $\epsilon^{R,\mathrm{NSI}}$, 
single-parameter bounds on the off-diagonal parameters are given in \autoref{tab:bounds}.
}
\label{fig:exclusion}
\end{figure}

Considering two parameters non-vanishing at the same time, we obtain the exclusion plots given in \autoref{fig:exclusion}. Note that here we do not show $\epsilon^P_\mu$ (recall the definition in \eqref{eq:epseffective}), since all the plots and bounds are indistinguishable from using $\epsilon^S_\mu$. In the case of $\epsilon^{S}_\mu$ or $\epsilon^{P}_\mu$ and $\epsilon^T_\mu$ non-zero, we assume real parameters for simplicity. It is worth noting that in the $L$-$R$-case introducing a flux uncertainty rotates the elliptical shape of the allowed region. This can be understood considering the cross sections \eqref{eq:diffcross}, when only 
$\epsilon^{L,\mathrm{NSI}}_{\mu\mu}$ and $\epsilon^{R,\mathrm{NSI}}_{\mu\mu}$ are introduced, such that we essentially employ the Standard Model coefficients \eqref{eq:abcdSM} and vary $g_L$ and $g_R$. For instance, the neutrino channel differential cross section would read
\[
\frac{\der\sigma_{\nu_\mu\rightarrow\nu_\beta}}{\der T} =
\frac{2G_F^2m_e}{\pi}
\left[
g_L^2+g_R^2\left(1-\frac{T}{E_\nu}\right)^2 - g_Lg_R\frac{m_eT}{E_\nu^2}
\right].
\]
Now since $g_L$ is negative while $g_R$ is positive, slightly increasing (decreasing) one can be approximately compensated for by increasing (decreasing) the other, which explains the direction of the ellipsis when flux normalisation is neglected (red contour in the top-left corner of \autoref{fig:exclusion}). On the other hand, increasing (decreasing) one while decreasing (increasing) the other can be approximately compensated for by the normalisation parameter $a$, which explains the opposite direction of the blue and green ellipses. This demonstrates the importance of a high-precision flux determination for the exclusion of (approximately) degenerate directions in parameter space. \\

Let us discuss if one could solidly distinguish the different new physics effects. To this end we consider the spectra generated by one parameter tuned away from the SM and fit the result to the other parameters (assuming configuration B defined in \eqref{eq:configurations}). The result is what one could expect from \eqref{eq:diffcross} and \eqref{eq:abcd}, together with \eqref{eq:lrsplit}: There are degeneracies between $\epsilon^{L,\mathrm{NSI}}_{\mu\mu}$ and $\epsilon^{L,\mathrm{NSI}}_{\mu e},$ $\epsilon^{L,\mathrm{NSI}}_{\mu\tau}$ in negative direction of $\epsilon^{L,\mathrm{NSI}}_{\mu\mu}$, but positive $\epsilon^{L,\mathrm{NSI}}_{\mu\mu}$ could be well distinguished. On the other hand, there is a degeneracy between $\epsilon^{R,\mathrm{NSI}}_{\mu\mu}$ and $\epsilon^{R,\mathrm{NSI}}_{\mu e},$ $\epsilon^{R,\mathrm{NSI}}_{\mu\tau}$ in positive direction of $\epsilon^{R,\mathrm{NSI}}_{\mu\mu}$, but negative $\epsilon^{R,\mathrm{NSI}}_{\mu\mu}$ could be well distinguished. The difference in sign stems from the fact that $g_L$ is negative while $g_R$ is positive. Moreover, there is an obvious degeneracy between $\epsilon^S_\mu$ and $\epsilon^P_\mu$ which one cannot hope to distinguish. Finally, $\epsilon^T_\mu$-effects can be distinguished from the others. The discriminative power clearly depends on the magnitude of new interactions, but let us give one example. If we assume a spectrum is generated with $\epsilon^{S}_\mu=0.2$, then the best-fit of the other parameters (except for $\epsilon^P_\mu$) leads to $\chi^2\geq17$, or even $\chi^2\geq20$ if flux uncertainty is negligible. We conclude that both exotic new physics effects and NSI have the potential to be identified distinctively at DUNE ND.

Now we briefly discuss the advantages of the differential cross section analysis and thereby explain how our analysis extends the discussion in \cite{falkowski}. Therein, a longer exposure time of 3+3 years and a larger detector of 100 tons is assumed, which leads to higher, but comparable statistics. Expected constraints on $\epsilon^{L,\mathrm{NSI}}_{\mu\mu}$ and $\epsilon^{R,\mathrm{NSI}}_{\mu\mu}$ are calculated via the ratio $R$ of total event numbers and the SM prediction, and given in terms of a two-parameter plot, such that we cannot compare explicit numerical values. Assuming statistically dominated errors, we obtain a plot, represented by the red contour in the top-left corner of \autoref{fig:exclusion}, that is very similar in shape and order of magnitude to Figure 1 in \cite{falkowski}.
It is interesting to note, however, that such ratios $R$ are directly sensitive to global systematics like flux uncertainties. Indeed it is found that assuming a 1\% systematic error on the measurements of $R$ leads to significantly weaker bounds than assuming statistics to be dominating. In our case, due to the use of spectral information, the 1\% uncertainty has a much less significant effect on the constraints in \autoref{tab:bounds} and \autoref{fig:exclusion}. Therefore, one major advantage is that the method is less sensitive to important experimental systematics.\footnote{Here we have only considered flux uncertainties, but other, in particular energy-dependent systematics could readily be included in the actual fit.} 
The second major advantage, as we already discussed, is that different new physics effects can potentially be distinguished much better. This is because an increase of total cross sections can, in principle, originate from any of the parameters, while their effect on differential cross sections will be distinct.

\section{Conclusion}
\label{sec:conclusion}
We have investigated prospects of the planned DUNE near-detector facility to find or constrain new physics in the lepton sector. Namely, from neutrino-electron scattering events, one can expect to probe general effective $\nu\nu ee$-couplings with up to an order of magnitude improved precision (according to \autoref{tab:bounds}). In particular, the bounds on flavor-diagonal Non-Standard Interactions, $\epsilon^{L,\mathrm{NSI}(e)}_{\mu\mu}$, and $\epsilon^{R,\mathrm{NSI}(e)}_{\mu\mu}$ can be probed down to the order $10^{-3}$ after the first five years of operation. This implies that for some couplings, the measurement will allow to probe effective mass scales of up to \SI{9}{TeV} assuming statistically dominated uncertainties (\SI{7}{TeV} assuming 1\% flux normalisation uncertainty, or \SI{6}{TeV} assuming 1\% flux normalisation uncertainty). We have demonstrated that, if new physics is found, one can potentially distinguish Non-Standard Interactions from exotic (pseudo)scalar and tensor interactions. Moreover, we pointed out that bounds from scatterings at the near detector are competitive, and in parts complementary, to the bounds from oscillation data. They may even be a key input to the reliable measurement of the leptonic CP phase by constraining the possible interference of Non-Standard Interactions.

It will be interesting to investigate if the inclusion of bounds on new neutrino-quark interactions from scattering data can be combined to sufficient bounds on Non-Standard Interactions in matter to notably improve the sensitivity of oscillation experiments towards the fundamental parameters of neutrino physics.
Neutrino-electron scattering will in the future also be studied by other experiments, which will allow for further exploration of the parameter spaces. In addition, other processes like coherent neutrino-nucleus scattering can also probe general neutrino interactions, which still remains a field with few analyses.

\subsection*{Acknowledgements}
We thank Xun-Jie Xu for helpful discussions. 
IB is supported by the IMPRS-PTFS and enrolled at Heidelberg University. WR is supported by the DFG with grant RO 2516/7-1 in the Heisenberg program.

\appendix

\section{Relation to SMEFT operators}
\label{sec:smeft}
In this section, we discuss the possible origin of new physics, parametrised in terms of \eqref{eq:lagrangian} from Standard Model effective field theory (SMEFT) operators. In SMEFT, one considers all gauge invariant operators built from SM fields up to a certain mass dimension and expands the effective Lagrangian in those. To obtain terms of the correct mass dimension, one typically multiplies by factors of the inverse of some new-physics mass scale $\Lambda$, which is interpreted as the cutoff of the EFT,
\[
\lagr_{\mathrm{eff}} = \lagr_{\mathrm{SM}}+\sum_{n\geq5}\frac{1}{\Lambda^{n-4}}\mathcal{O}^{(n)}\,. 
\]
Restricting ourselves to the dimension-six case, instead of $1/\Lambda^2$, we multiply operators by $\sqrt{8}G_F$ and hence compare dimensionless operator coefficients with the weak scale in the spirit of \eqref{eq:lagrangian}.

At the dimension-six level, a non-redundant set of operators is the Warsaw basis \cite{warsaw}. 
The relevant four-fermion operators for generating $\epsilon^L$ and $\epsilon^R$ in this basis are
\begin{align}
\mathcal{O}_{\ell \ell} &= C^{\ell\ell}_{\alpha\beta\gamma\delta} (\overline \ell_\alpha \gamma_\mu \ell_\beta)(\overline \ell_\gamma \gamma^\mu \ell_\delta)\,,\label{eq:SMEFT1}\\
\mathcal{O}_{\ell e} &= C^{\ell e}_{\alpha\beta\gamma\delta} (\overline \ell_\alpha \gamma_\mu \ell_\beta)(\overline e_\gamma \gamma^\mu e_\delta)\,,\label{eq:SMEFT2}
\end{align}
where $\ell_\alpha$ denotes a left-handed lepton doublet of flavor $\alpha$, $e_\alpha$ denotes a right-handed (charged) lepton singlet of flavor $\alpha$. The Wilson coefficients $C^{\ell\ell}_{\mu\beta ee}$ and $C^{\ell e}_{\mu\beta ee}$ then generate NSI parameters
\[
\epsilon^{L,\mathrm{NSI}}_{\mu\beta} = 4C^{\ell\ell}_{\mu\beta ee} + 4C^{\ell\ell}_{ee\mu\beta}\,, \qquad
\epsilon^{R,\mathrm{NSI}}_{\mu\beta} = 4C^{\ell e}_{\mu\beta ee}\,.
\]
However, those operators also give rise to corrections to scatterings of four charged leptons of equal magnitude and are therefore further constrained by electro-weak precision data. Another way to generate NSI is by the scalar-fermion operators which correct the $W$ and $Z$ couplings of fermionic currents after electro-weak symmetry breaking, namely $\mathcal{O}_{\varphi\ell}^{(1)}$, $\mathcal{O}_{\varphi\ell}^{(3)}$, and $\mathcal{O}_{\varphi e}$ in \cite{warsaw}, though we will not go into details here. 

It is more intricate to generate the (pseudo)scalar and tensor interactions. In fact all dimension-6 operators involving four leptons, including two neutrinos can be reformulated in terms of \eqref{eq:SMEFT1} and \eqref{eq:SMEFT2} \cite{warsaw}. Thus, in the Majorana case, no (pseudo)scalar or tensor interactions can be generated at dimension-six level without adding further fields.
In the Dirac case, where we add a right-handed singlet neutrino $N_\alpha$, the possible additional dimension-six operators have been studied in \cite{liao}. Many of those generate right-handed NSI $\widetilde\epsilon^L$, $\widetilde\epsilon^R$. We only reproduce those giving rise to (pseudo)scalar or tensor interactions,
\begin{align}\label{eq:NSMEFT1}
\mathcal{O}_{\varphi Ne} &= C^{\varphi Ne}_{\alpha\beta}i\left(\widetilde{\varphi}^\dagger D_\mu \varphi\right)
\left(\overline N_\alpha\gamma^\mu e_\beta\right)+\mathrm{h.c.},\\\label{eq:NSMEFT2}
\mathcal{O}_{\ell N\ell e} &= C^{\ell N\ell e}_{\alpha\beta\gamma\delta}\left(\overline{L}^i_\alpha N_\beta\right)\epsilon_{ij}
\left(\overline L^j_\gamma e_\delta\right)+\mathrm{h.c.}
\end{align}
where $\varphi$ and $\widetilde\varphi$ denote the Higgs doublet and its conjugate, while $\epsilon_{ij}$ denotes the Levi-Civita symbol with $SU(2)$ indices $i,j$ which are summed over. The first operator gives rise to a right-handed leptonic charged current coupling to $W$, and hence, by integrating out a $W$ boson propagator, effective four-fermion vertices would arise in analogy to Fermi theory. Fierz transforming those into the neutral-current form, (pseudo)scalar interactions can be generated, namely
\[
\epsilon^S_{\mu e} = -\epsilon^P_{\mu e} = \widetilde\epsilon^{S*}_{e \mu} = \widetilde\epsilon^{P*}_{e \mu} = 2C^{HN\ell}_{\mu e}\,.
\]
The second operator directly gives rise to effective four-fermion vertices. We find
\begin{align}
\epsilon^S_{\mu\beta}=\epsilon^P_{\mu\beta}
=\widetilde\epsilon^{S*}_{\beta\mu} &= -\widetilde\epsilon^{P*}_{\beta\mu}
=\left(C^{\ell N\ell e}_{\beta\mu ee}\right)^*+\frac12\left(C^{\ell N\ell e}_{e\mu\beta e}\right)^*,\\
\epsilon^T_{\mu\beta}&=\widetilde\epsilon^{T*}_{\beta\mu} = \frac18\left(C^{\ell N\ell e}_{e\mu\beta e}\right)^*,
\end{align}
such that (pseudo)scalar and tensor interactions are generated simultaneously. We leave details on the derivation of the coefficients from \eqref{eq:NSMEFT1} and \eqref{eq:NSMEFT2} for future work.

\section{Mapping of parametrisations}
\label{sec:CDparametrisation}
Here we discuss the mapping between the parametrisation of general neutrino-electron interactions in \eqref{eq:lagrangian}, named the $\epsilon$-parametrisation, and the one employed in e.g. \cite{Kayser:1979mj,Rosen:1982pj,xu}, which we refer to as the $CD$-parametrisation. The equivalent to \eqref{eq:lagrangian} reads\footnote{We employ the convention of \cite{xu} which differs from the one in \cite{Kayser:1979mj,Rosen:1982pj} only by factors of the imaginary unit $i$.}
\begin{align}
\lagr&=-\frac{G_F}{\sqrt{2}}\sum_{a=S,P,V,A,T}\left(\overline{\nu}_\alpha\,\Gamma^a\nu_\beta\right)
\left(\overline{e}\Gamma^a(C^a_{\alpha\beta}+\overline{D}^a_{\alpha\beta}i\gamma^5)e\right),
\end{align}
where the five possible independent combinations of Dirac matrices $\Gamma^a$ are defined as
\[
\Gamma^a\in \left\{\mathds{1},i\gamma^5,\gamma^\mu,\gamma^\mu\gamma^5,\sigma^{\mu\nu}\right\},
\]
and the coefficients $C^a_{\alpha\beta}$ and 
\[
D^a_{\alpha\beta}\equiv\begin{cases}
\overline{D}^a_{\alpha\beta} & (a=S,P,T) \\
i\overline{D}^a_{\alpha\beta}& (a=V,A)\end{cases}
\]
obey
\[\label{eq:CDrelation}
C^a_{\alpha\beta} = C^{a*}_{\beta\alpha}\,,\quad D^a_{\alpha\beta} = D^{a*}_{\beta\alpha}\,,
\]
namely, they form hermitian matrices in flavor space. 
The complete relations between the parametrisations $(\epsilon,\widetilde\epsilon)\,\leftrightarrow\,(C,D)$ read
\[
\begin{split}
\epsilon^L &= \frac14\left(C_V-D_V+C_A-D_A\right),\\
\epsilon^R &= \frac14\left(C_V+D_V-C_A-D_A\right),\\
\epsilon^S&=\frac12\left(C_S+iD_P\right),\\
-\epsilon^P&=\frac12\left(C_P+iD_S\right),\\
\epsilon^T&=\frac14\left(C_T-iD_T\right),
\end{split}
\qquad
\begin{split}
\widetilde\epsilon^L &= \frac14\left(C_V-D_V-C_A+D_A\right),\\
\widetilde\epsilon^R &= \frac14\left(C_V+D_V+C_A+D_A\right),\\
\widetilde\epsilon^S&=\frac12\left(C_S-iD_P\right),\\
-\widetilde\epsilon^P&=\frac12\left(-C_P+iD_S\right),\\
\widetilde\epsilon^T&=\frac14\left(C_T+iD_T\right),
\end{split}
\]
where flavor indices are suppressed.

\section{Derivation of the cross sections}
\label{sec:derivation}
Here we derive the differential cross sections of neutrino-electron scattering for a generic interaction Lagrangian \eqref{eq:lagrangian}.
We consider the ingoing and outgoing (anti)neutrinos massless. The associated Feynman-rule quantities are $u_-(p_\nu)$ (and $\overline{v_+}(p_\nu)$ respectively) for the ingoing (anti)neutrino and $\overline{u_s}(k_\nu)$ (and $v_s(k_\nu)$) for the outgoing (anti)neutrino of helicity $s=\pm$.
We organised the new interactions in terms of chiral operators. Hence, it is easy to see that the operators including one gamma matrix ($V/A$ or $j=1,2,3,4$) admit processes $\nu_{\alpha,L}\rightarrow \nu_{\beta,L}$ or $\overline{\nu}_{\alpha,R}\rightarrow \overline{\nu}_{\beta,R}$, while the operators containing no or two gamma matrices ($S,P,T$ or $j=5,\dots,10$) mediate chirality flipping processes $\nu_{\alpha,L}\rightarrow \nu_{\beta,R}$ or $\overline{\nu}_{\alpha,R}\rightarrow \overline\nu_{\beta,L}$.

The amplitudes of the scattering $\nu_\alpha+e^{-}_r\rightarrow \nu_{\beta,s}+e^{-}_{r'}$, and $\overline\nu_\alpha+e^{-}_r\rightarrow \nu_{\beta,s}+e^{-}_{r'}$ read
\begin{align}
\mathcal{A}_{\alpha\beta}^{srr'}&=\frac{G_F}{\sqrt{2}}\sum_{j=1,3,5,7,9}
{\epsilon}^j_{\alpha\beta}
\left(\overline{u_s}(k_{\nu})\mathcal{O}_ju_-(p_\nu)\right)
\left(\overline{u_{r'}}(k_e)\mathcal{O}_j'u_r(p_e)\right),\\
\mathcal{A}_{\overline\alpha \overline\beta}^{srr'}&=\frac{G_F}{\sqrt{2}}\sum_{j=1,3,6,8,10}
\epsilon^j_{\beta\alpha}
\left(\overline{v_+}(p_\nu)\mathcal{O}_jv_s(k_{\nu})\right)
\left(\overline{u_{r'}}(k_e)\mathcal{O}_j'u_r(p_e)\right).
\end{align}
To obtain cross sections we compute
\begin{align}
|\mathcal{A}_{\alpha\beta}|^2&=\frac12\sum_{s,r,r'}|\mathcal{A}^{srr'}_{\alpha\beta}|^2\,,\\
|\mathcal{A}_{\overline\alpha\overline\beta}|^2&=\frac12\sum_{s,r,r'}|\mathcal{A}^{srr'}_{\overline\alpha\overline\beta}|^2\,,
\end{align}
where the factor of $1/2$ stems from averaging over initial electron helicity. Although the helicity $s$ of the outgoing (anti)neutrino is fixed by each specific operator, it is convenient to sum over both helicities to apply trace identities when calculating the square-amplitude. The differential cross sections are then obtained by the standard formula
\[
\frac{\der\sigma}{\der T}=\frac{|\mathcal{A}|^2}{32\pi m_eE_\nu^2}\,.
\]
Before arriving at the form of \eqref{eq:diffcross}, we need to apply \eqref{eq:epsilonconstraints} to substitute expressions like 
\[
|\epsilon^{L/R}_{\beta\alpha}|^2\,\, \rightarrow\,\, |\epsilon^{L/R}_{\alpha\beta}|^2 \,,\quad
|\widetilde\epsilon^{S/P/T}_{\beta\alpha}|^2 \,\,\rightarrow\,\, |\epsilon^{S/P/T}_{\alpha\beta}|^2\,,\quad\dots\,,
\]
thereby reducing the expression to the minimum number of free parameters.

\bibliographystyle{utphys}
\bibliography{Dune-constraints}{}
\end{document}